\documentstyle[prl,aps,preprint,tighten,epsfig]{revtex}
\begin{document}
\draft
\title{Extension of the Cosmic-Ray Energy Spectrum \\
Beyond the Predicted Greisen-Zatsepin-Kuz'min Cutoff
}

\author{
{M. Takeda$^{1}$,}
{N. Hayashida$^{1}$,} 
{K. Honda$^{2}$,}
{N. Inoue$^{3}$,}
{K. Kadota$^{4}$,}
{F. Kakimoto$^{4}$,}	\\
{K. Kamata$^{5}$,}
{S. Kawaguchi$^{6}$,}
{Y. Kawasaki$^{7}$,}
{N. Kawasumi$^{8}$,}
{H. Kitamura$^{9}$,}	\\
{E. Kusano$^{3}$,}
{Y. Matsubara$^{10}$,}
{K. Murakami$^{11}$,}
{M. Nagano$^{12}$,}
{D. Nishikawa$^{1}$,}	\\
{H. Ohoka$^{1}$,}
{N. Sakaki$^{1}$,}
{M. Sasaki$^{1}$,}
{K. Shinozaki$^{3}$,}
{N. Souma$^{3}$,}
{M. Teshima$^{1}$,}	\\
{R. Torii$^{1}$,}
{I. Tsushima$^{8}$,}
{Y. Uchihori$^{13}$,}
{T. Yamamoto$^{1}$,}
{S. Yoshida$^{1}$}
{and H. Yoshii$^{14}$}	\\
{\hfill}
}

\address{$^{1}$ 
Institute for Cosmic Ray Research, University of Tokyo, Tokyo 188-8502, Japan
}
\address{$^{2}$ 
Faculty of Engineering, Yamanashi University, Kofu 400-8511, Japan
}
\address{$^{3}$ 
Department of Physics, Saitama University, Urawa 338-8570, Japan
}
\address{$^{4}$ 
Department of Physics, Tokyo Institute of Technology, Tokyo 152-8551, Japan
}
\address{$^{5}$ 
Nishina Memorial Foundation, Komagome, Tokyo 113, Japan
}
\address{$^{6}$ 
Faculty of General Education, Hirosaki University, Hirosaki 036-8560, Japan
}
\address{$^{7}$ 
Department of Physics, Osaka City University, Osaka 558-8585, Japan
}
\address{$^{8}$ 
Faculty of Education, Yamanashi University, Kofu 400-8510, Japan
}
\address{$^{9}$ 
Department of Physics, Kobe University, Kobe 657-8501, Japan
}
\address{$^{10}$ 
Solar-Terrestrial Environment Laboratory, Nagoya University, 
Nagoya 464-8601, Japan
}
\address{$^{11}$ 
Nagoya University of Foreign Studies, Nissin, Aichi 470-0131, Japan
} 
\address{$^{12}$ 
11-401, 5-8 Higashi, Hasuda, Saitama 394-0111, Japan
}
\address{$^{13}$ 
National Institute of Radiological Sciences, Chiba 263-8555, Japan
}
\address{$^{14}$ 
Department of Physics, Ehime University, Matsuyama 790-8577, Japan
}
\maketitle
\begin{abstract}
The cosmic-ray energy spectrum above $ 10^{18.5} eV $ is reported 
using the updated data set of 
the Akeno Giant Air Shower Array (AGASA) 
from February 1990 to October 1997. 
The energy spectrum extends beyond $ 10^{20} eV $ and 
the energy gap between the highest energy event and the others 
is being filled up with recently observed events. 
The spectral shape suggests 
the absence of the $ 2.7 \, K $ cutoff in the energy spectrum 
or a possible presence of a new component 
beyond the $ 2.7 \, K $ cutoff. 
\end{abstract}
\medskip


How high the maximum energy of cosmic rays reaches is 
one of the most important problems in cosmic ray research. 
Detections of cosmic rays 
with energies above $ 10^{20} eV $ \cite{bird95a,hayashida94a} 
have given rise to much discussion regarding their origin. 
Many models have been proposed as source candidates 
of such high energy cosmic rays: 
active astrophysical objects \cite{mdl_object}, 
decay products of much higher energy particles such as 
superheavy relic particles \cite{mdl_relic}
or 
topological defects \cite{mdl_td}, 
or cosmological gamma-ray bursts \cite{mdl_grb}
(see Ref.\ \cite{berezinsky98a} for a recent review). 
If such high energy cosmic rays come from far outside our Galaxy, 
they interact with cosmic microwave background photons 
and cannot travel cosmological distances. 
This interaction causes a cutoff in the energy spectrum 
near $ 5 \times 10^{19} eV $ 
which is referred to as 
the Greisen-Zatsepin-Kuz'min (GZK) cutoff \cite{gzk66}. 
Furthermore, the cosmic rays which have interacted form 
a ``bump'' just below the GZK cutoff energy 
\cite{hill85a,berezinsky88a,yoshida93a}. 
The change in the spectral slope around $ 10^{19} eV $ (``ankle'') may 
arise from a transition 
from galactic to extragalactic sources. 
The investigation of these features in the energy spectrum 
is one of the most important scientific challenges. 

There are two techniques for detecting extensive air showers (EAS): 
widely spread surface arrays and 
atmospheric fluorescence detectors. 
Using these techniques, 
the energy spectrum of extremely high energy cosmic rays 
has been measured by many groups such as 
Volcano Ranch \cite{linsley86a}, 
Haverah Park \cite{lawrence91a}, 
Sugar \cite{winn86a}, 
Yakutsk \cite{efimov91a}, 
Fly's Eye \cite{bird94a}, 
and Akeno \cite{nagano92a,yoshida95a}
(only the Fly's Eye group has adopted 
the atmospheric fluorescence detector). 
While the energy spectrum obtained from these experiments 
coincide within $ \pm 15 \% $ in energy below $ \sim 10^{19} eV $, 
the details of energy spectrum in the highest energy range 
is still inconclusive, mainly because of low statistics of 
their observed events. 
In this letter, 
we present the energy spectrum above $ 10^{18.5} eV $ obtained from 
the Akeno Giant Air Shower Array (AGASA) \cite{chiba92a,ohoka97a}, 
which currently has the largest exposure of 
any extremely high energy cosmic ray detectors.

The AGASA array is the largest operating surface array, 
covering an area of about $ 100 \, km^{2} $ and 
consisting of $ 111 $ surface detectors of $ 2.2 \, m^{2} $ area. 
Each surface detector is placed with a nearest-neighbor separation 
of about $ 1 \, km $ and 
the detectors are sequentially connected with pairs of optical fibers. 
All the detectors are controlled at detector sites 
through rapid communication with a central computer. 
The data acquisition system of AGASA was improved 
in December 1995 \cite{ohoka97a}. 
In a widely spread surface array like AGASA, 
the local density of charged shower particles 
at a specific distance from the shower axis 
is well established as an energy estimator 
\cite{hillas71a}, 
since this depends weakly on variation in the interaction model, 
fluctuation in shower development and the primary mass. 
In the AGASA experiment, 
we adopt local density $ S(600) $ at $ 600 \, m $ 
which is determined from fitting 
the lateral distribution of observed particle densities 
to an empirical formula \cite{yoshida94a}. 
This empirical formula is found to be valid for 
EAS with energies up to $ 10^{20} eV $ \cite{doi95a,sakaki98a}. 
The conversion relation from $ S(600) $ to the primary energy 
is evaluated through the Monte Carlo simulation \cite{dai88a} 
up to $ 10^{19} eV $ by 
\begin{displaymath}
	E = 2.03 \times 10^{17} \, S_{0}(600) \hspace{1em} eV, 
\end{displaymath}
where $ S_{0}(600) $ is the $ S(600) $ value in units of $ m^{-2} $ 
for a vertically incident shower. 
Since an inclined air shower traverses more atmospheric depth 
than a vertical shower, 
$ S_{\theta}(600) $ observed with zenith angle $ \theta $ must be 
transformed into $ S_{0}(600) $ at the vertical. 
This attenuation curve of $ S(600) $ has been formulated 
by Yoshida {\it et al} \cite{yoshida94a}. 

\begin{figure}
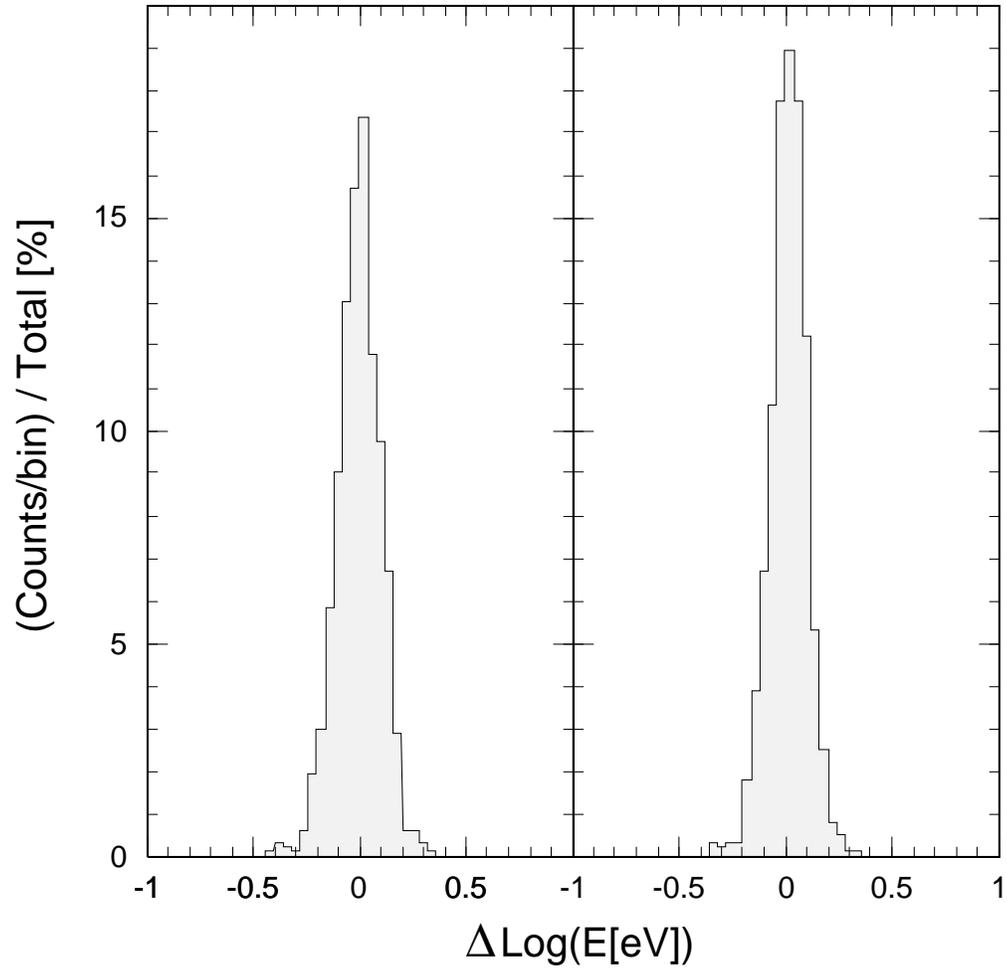

\caption{Fluctuation of energy determination 
	for $ 10^{19.5} eV $ (left) and $ 10^{20} eV $ (right) showers 
	with zenith angles less than $ 45^{\circ} $.}
\label{fig:fluct100EeV}
\end{figure}

The accuracy of event reconstruction has been evaluated 
through the analysis of a large number of artificial air shower events. 
These artificial events were simulated 
over a larger area than the AGASA area 
with directions sampled from an isotropic distribution. 
In this air shower simulation, 
the fluctuation on the longitudinal development of air showers, 
the resolution of the scintillation detectors, 
and statistical fluctuation of observed shower particles 
at each surface detector 
were taken into account. 
Only events with zenith angles smaller than $ 45^{\circ} $ and 
with core locations inside the array area are 
used in the following analysis. 
Fig.\ \ref{fig:fluct100EeV} shows the fluctuation of energy determination 
for $ 10^{19.5} eV $ (left) and $ 10^{20} eV $ (right) showers 
with zenith angles less than $ 45^{\circ} $. 
The primary energy is determined with an accuracy of 
about $ \pm 30 \% $ 
and the proportion of events with a $ 50 \% $-or-more overestimation 
in energy is about $ 2.4 \% $. 

Energy uncertainty also arises from the following systematic errors. 
The first is uncertainty in measuring the particle density 
incident upon each detector. 
The number of incident particles is determined from 
the time width of a pulse, 
which is generated by decaying an anode signal 
of a photomultiplier tube 
exponentially with a time constant of about $ 10 \mu s $ 
and discriminated at a certain level 
(see \cite{chiba92a} for the details of the AGASA instruments). 
The variation in the amplifier gain and the decay constant 
are monitored in every run for detector calibration 
and their seasonal variations are within $ 2 \% $. 
The second is uncertainty in 
the empirical formula of the lateral distribution function and 
in the attenuation curve of $ S(600) $. 
The energy uncertainty due to the limited accuracy on 
both of these is estimated to be $ \pm 20 \% $, 
even if both factors shift the estimated energy 
in the same direction \cite{yoshida94a}. 
The third is uncertainty in the conversion formula of $ S(600) $ 
into primary energy. 
Although this formula is not sensitive to 
interaction models or primary composition 
in each of simulation codes \cite{dai88a}, 
the systematic errors due to the differences in simulation codes 
are not quantitatively clear. 

In order to evaluate the systematic errors {\em experimentally}, 
we compare the AGASA spectrum derived below 
with the Akeno spectrum 
which was accurately determined 
between $ 10^{14.5} eV $ and $ 10^{19} eV $ 
using the arrays with different detector spacing \cite{nagano92a}. 
The Akeno spectrum fits very well with extrapolation of 
those obtained from direct measurement on balloons and satellites, 
and with the Tibet result \cite{tibet96a} 
obtained through the observation of the shower 
at the height of its maximum development. 
The difference between the present AGASA and Akeno spectra is 
about $ 10 \% $ in energy at $ 10^{18.5} eV $. 
In addition, 
the difference among spectra obtained from 
the Fly's Eye, Yakutsk, Haverah Park, 
and AGASA experiments 
is within $ 30 \% $ in energy 
in spite of quite different methods 
for determining the primary energy. 
Therefore, the total systematic error in the AGASA energy estimation 
is estimated to be within $ 30 \% $, 
and the primary energy of the highest energy event of AGASA, 
for example, 
is estimated to be in the range ($1.7$ -- $2.0$) $ \times 10^{20} eV $.

The effective area of AGASA has been calculated from 
the simulation of artificial air shower events. 
The energy spectrum in this simulation was assumed to be $ E^{-3} $, 
and the reconstruction uncertainty in energy estimation was also 
taken into account. 
Although the effective area depends weakly on the spectral index, 
this dependence is negligible when compared with other ambiguities
like energy resolution. 
The total exposure of AGASA 
is obtained by multiplying the effective area and 
the observation time of each branch for each epoch. 
Above $ 10^{19} eV $, 
this exposure is constant and is $ 2.6 \times 10^{16} \, m^{2} \, sr \, s $, 
which is about five times as large as 
that in our previous paper \cite{yoshida95a} 
(cf. $ \sim 0.5 \times 10^{16} \, m^{2} \, sr \, s $ 
of the stereo Fly's Eye exposure \cite{bird94a}
and  $ \sim 0.7 \times 10^{16} \, m^{2} \, sr \, s $ 
of the Haverah Park exposure \cite{lawrence91a}). 
However, 
the exposure below $ 10^{18.5} eV $ depends strongly 
on the primary energy. 
Since this energy dependence causes systematic errors 
in the energy spectrum derivation, 
only events with energies above $ 10^{18.5} eV $ 
are used for the energy spectrum in this letter. 
From February 1990 to October 1997, 
$ 3847 $, $ 461 $ and $ 6 $ events were observed 
with energies above $ 10^{18.5} eV $, $ 10^{19} eV $ 
and $ 10^{20} eV $ respectively.

\begin{figure}
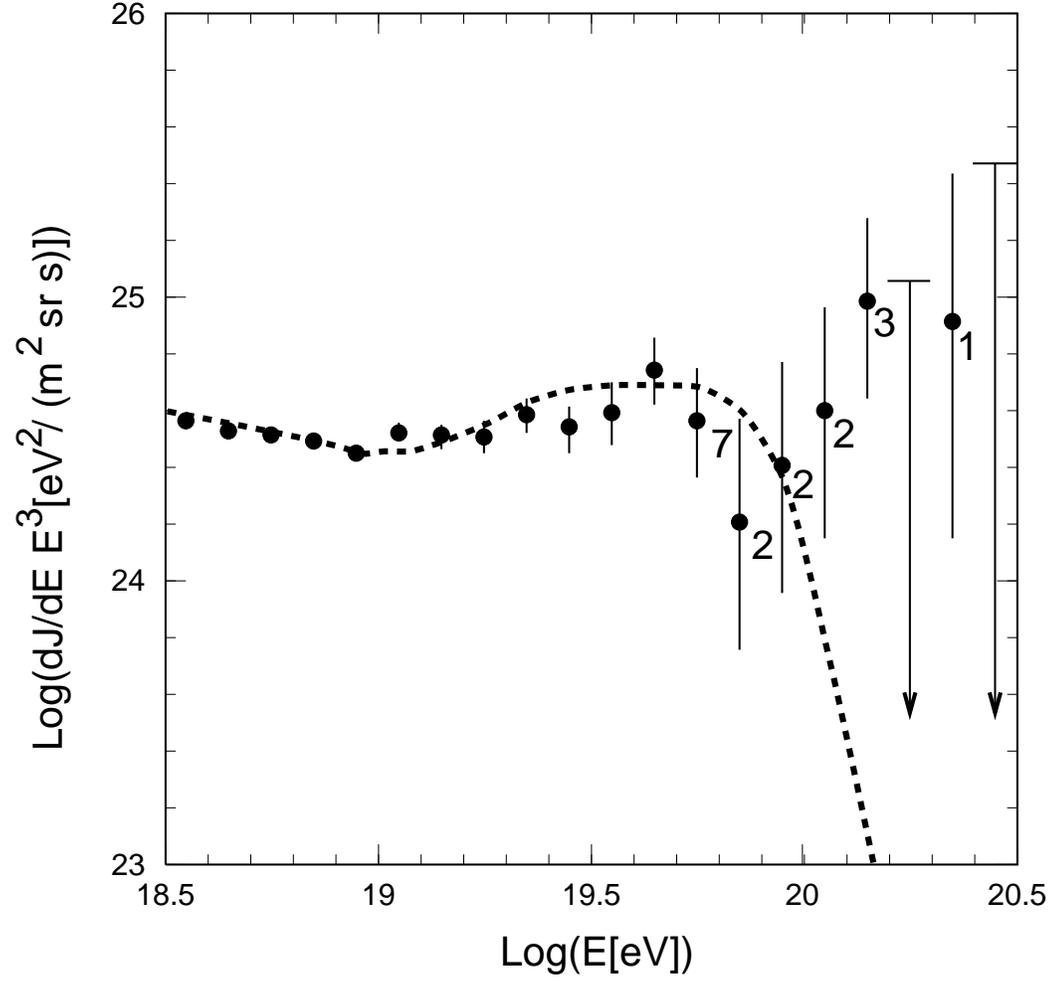

\caption{Energy spectrum observed with AGASA.
	The vertical axis is multiplied by $ E^{3} $. 
	Error bars represent the Poisson upper and lower limits 
	at $ 68 \% $ and arrows are $ 90 \% $ C.L. upper limits. 
	Numbers attached to points show the number of events 
	in each energy bin.
	The dashed curve represents the spectrum expected for 
	extragalactic sources distributed uniformly in the Universe, 
        taking account of the energy determination error [11].}
\label{fig:energy_spectrum}
\end{figure}

The energy spectrum observed with AGASA is shown 
in Fig.\ \ref{fig:energy_spectrum}, 
multiplied by $ E^{3} $ in order to emphasize details of 
the steeply falling spectrum. 
Error bars represent the Poisson upper and lower limits at $ 68 \% $ 
and arrows are $ 90 \% $ C.L. upper limits. 
Numbers attached to points show the number of events in each energy bin. 
The dashed curve represents the spectrum expected for 
extragalactic sources distributed uniformly in the Universe, 
taking account of the energy determination error \cite{yoshida93a}. 

First, we examine whether the observed energy spectrum could be 
represented by a single power law spectrum ($ \propto E^{-\gamma_{1}} $). 
The optimum spectral index $ \gamma_{1} $ is derived from 
the maximum likelihood procedure comparing 
the observed and expected number of events in each energy bin. 
This procedure is same as described in Yoshida {\it et al.} 
\cite{yoshida95a}. 
%
The maximum likelihood procedure for a single power law spectrum 
results in $ \gamma_{1} = 3.08^{+0.08}_{-0.15} $; 
the likelihood significance of $ \gamma_{1} $ is only 0.051. 
If only events with energies below $ 10^{19} eV $ are considered, 
$ \gamma_{1}(E \leq 10^{19} eV) = 3.23^{+0.10}_{-0.12} $ is obtained 
which is consistent with the spectral index, $ 3.16 \pm 0.08 $, 
determined from the Akeno experiment \cite{nagano92a}. 

Next, a broken energy spectrum is examined 
with the same procedure. 
The broken energy spectrum is assumed to be 
\begin{displaymath}
	\frac{d J}{d E} = 
	\left\{ \begin{array}{lcc}
	\mbox{$ \kappa \left(E / E_{a} \right)^{-\gamma_{0}} $} 
	& & 10^{18.5} eV \leq E < E_{a} \\
	\mbox{$ \kappa \left(E / E_{a} \right)^{-\gamma_{2}} $} 
	& & E_{a} \leq E 
	\end{array} \right. ,
\end{displaymath}
where $ \gamma_{0} $ and $ \gamma_{2} $ are indexes below and above 
a bending (ankle) energy $ E_{a} $, and $ \gamma_{0} $ is fixed to 
be $ \gamma_{1}(E \leq 10^{19} eV) = 3.16 $ 
determined from the Akeno experiment \cite{nagano92a}. 
The most probable parameters are obtained 
at $ E_{a} = 10^{19.01} eV $ and $ \gamma_{2} = 2.78^{+0.25}_{-0.33} $, 
where the likelihood significance is found to be $ 0.903 $. 
This is also consistent with 
the results of $ 2.8 \pm 0.3 $ at energies above $ 10^{18.8} eV $ 
determined from the Akeno experiment \cite{nagano92a} 
and of $ 2.3^{+0.5}_{-0.3} $ above $ 10^{19.0} $ 
in the previous paper \cite{yoshida95a}. 

\begin{figure}
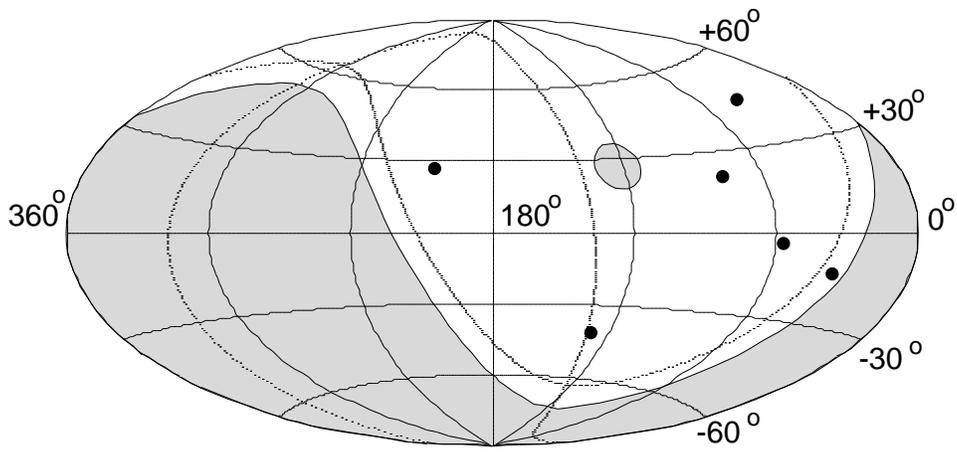

\caption{Arrival directions of six $ 10^{20} eV $ events 
	on the Galactic coordinates.
	The shaded regions indicate the non-observable celestial regions 
	due to the zenith angle cut of $ \leq 45^{\circ} $.
	The equatorial and supergalactic planes are also shown.}
\label{fig:directions}
\end{figure}

Furthermore, the energy spectrum presented here 
extends up to higher energies than 
the previous results \cite{nagano92a,yoshida95a}; 
six events were observed above $ 10^{20} eV $. 
If the real energy spectrum is that shown 
in Fig.\ \ref{fig:energy_spectrum} as the dashed curve, 
the expected number of events above $ 10^{20} eV $ is less than one, 
taking account of the energy resolution. 
The energy spectrum is therefore more likely to 
extend beyond $ 10^{20} eV $ without the GZK cutoff. 
However, it is also worth noting that 
the observed energy spectrum suggests 
a small deficit just below $ 10^{20} eV $, 
whose significance is not compelling 
because of the uncertainty in $ \gamma_{2} $ estimation. 
This deficit may imply another component above the GZK cutoff energy. 
In either case, 
sources of the most energetic cosmic rays 
must be located within a few tens of $ Mpc $ 
from our Galaxy \cite{yoshida93a}. 
The arrival directions of six $ 10^{20} eV $ events 
are shown in Fig. \ref{fig:directions}. 
Within the accuracy of arrival direction determination 
($ 1.6^{\circ} $ above $ 4 \times 10^{19} eV $), 
no $ 10^{20} eV $ events coincide with possible candidates from 
the second EGRET sources \cite{thompson95a} or 
the extragalactic radio sources 
with redshift $ z \leq 0.02 $ \cite{veron83a}. 
Our previous result for cosmic-ray arrival directions has been 
reported in Hayashida {\it et al.} \cite{hayashida96a} 
and the new results are under preparation. 

The fact that the energy spectrum 
extends beyond $ 10^{20} eV $ 
and no $ 10^{20} eV $ events coincide with 
nearby active astrophysical objects 
leads highest energy cosmic-ray physics 
into a much more exciting stage. 
The next generation experiments such as 
the Telescope Array \cite{teshima92a,teshima94a}, 
High Resolution Fly's Eye \cite{hires94a,alseady97a}, and 
Auger \cite{cronin92a,auger97a} projects 
will solve the puzzle of the highest energy cosmic rays. 

In conclusion, 
the cosmic-ray energy spectrum 
extends beyond $ 10^{20} eV $. 
No candidate sources are 
found in the directions of six $ 10^{20} eV $ events, 
while their sources must be closer than $ 50 \, Mpc $. 
The possible deficit around $ 10^{20} eV $ 
is a notable area 
in which to search for origin of the highest energy cosmic rays. 
Detailed discussion with the AGASA data will be published elsewhere.

We are grateful to Akeno-mura, Nirasaki-shi, Sudama-cho, Nagasaka-cho, 
Ohizumi-mura, Tokyo Electric Power Co. and 
Nihon Telegram and Telephone Co. for their kind cooperation. 
The authors are indebted to other members of the Akeno group 
in the maintenance of the AGASA array. 
The authors thank Jamie Holder and Michael Roberts for their 
valuable suggestion on the preparation of the manuscript.



\newcommand{\InsertFigure}[2]{\newpage\begin{center}\mbox{%
\epsfig{bbllx=1.4truecm,bblly=1.3truecm,bburx=19.5truecm,bbury=26.5truecm,%
height=21.4truecm,figure=#1}}\end{center}\vspace*{-1.8truecm}%
\parbox[t]{\hsize}{\small\baselineskip=0.5truecm\hspace*{0.5truecm} #2}}
\InsertFigure{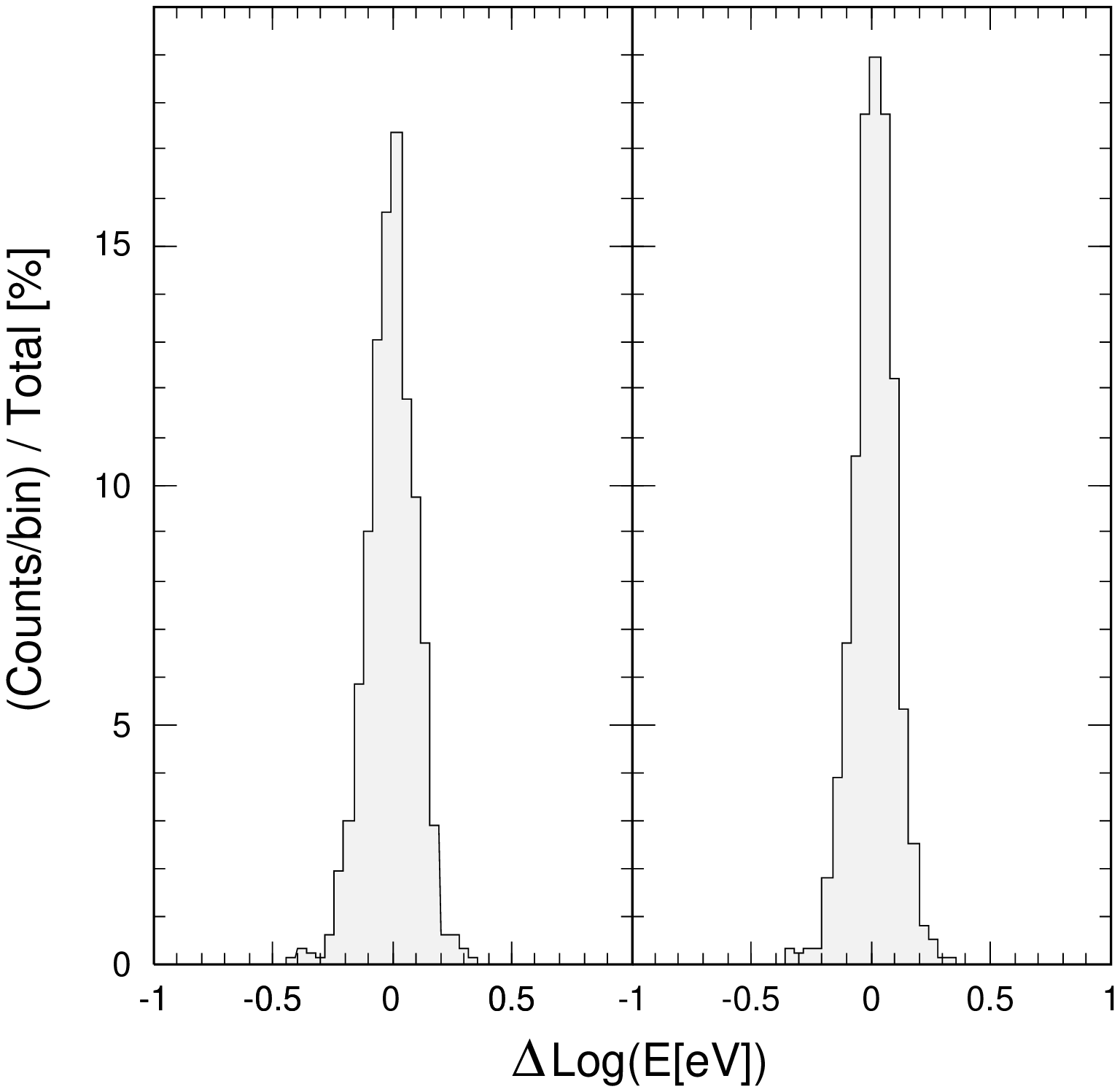}%
	{FIG.~1. Fluctuation of energy determination 
	for $ 10^{19.5} eV $ (left) and $ 10^{20} eV $ (right) showers 
	with zenith angles less than $ 45^{\circ} $.}

\InsertFigure{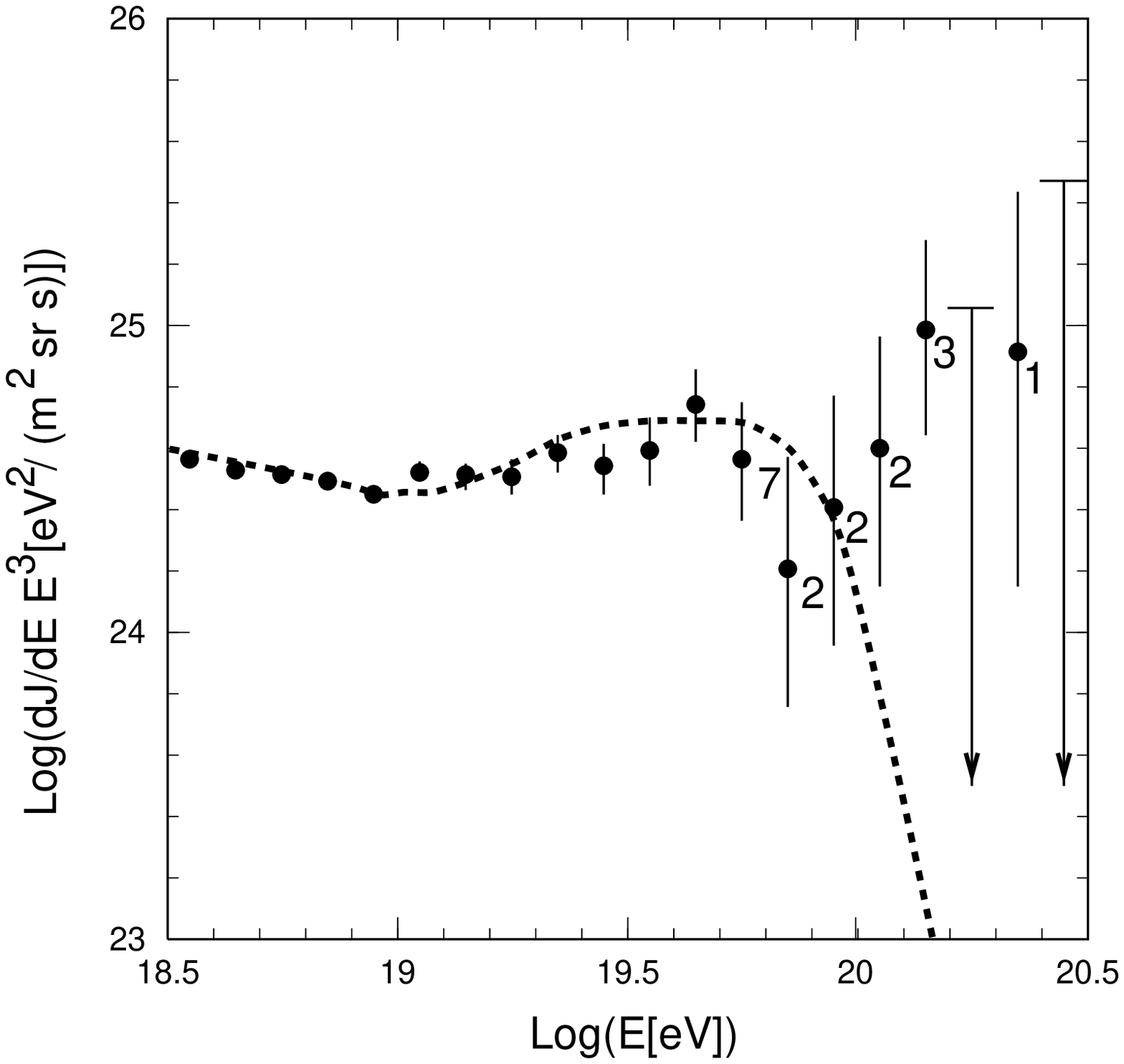}%
	{FIG.~2. Energy spectrum observed with AGASA.
	The vertical axis is multiplied by $ E^{3} $. 
	Error bars represent the Poisson upper and lower limits 
	at $ 68 \% $ and arrows are $ 90 \% $ C.L. upper limits. 
	Numbers attached to points show the number of events 
	in each energy bin.
	The dashed curve represents the spectrum expected for 
	extragalactic sources distributed uniformly in the Universe, 
        taking account of the energy determination error [11].}

\InsertFigure{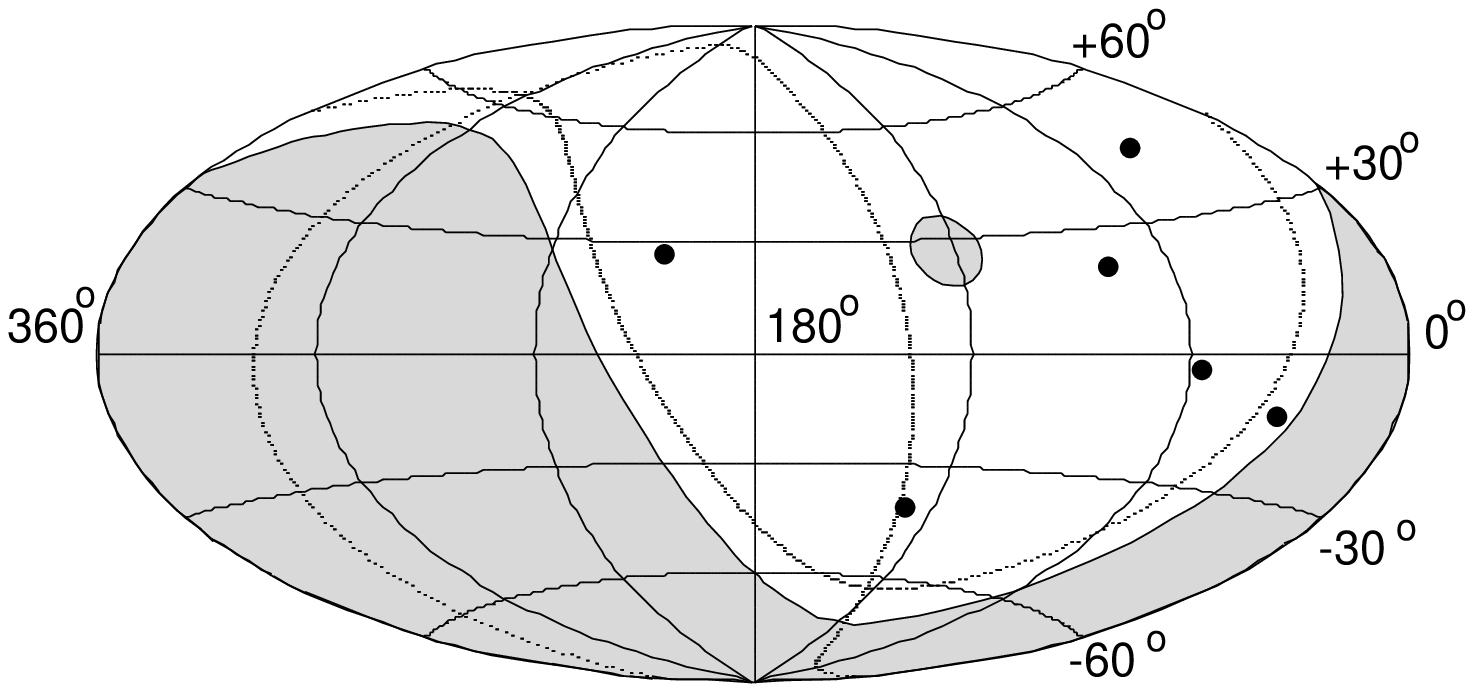}%
	{FIG.~3. Arrival directions of six $ 10^{20} eV $ events 
	on the Galactic coordinates.
	The shaded regions indicate the non-observable celestial regions 
	due to the zenith angle cut of $ \leq 45^{\circ} $.
	The equatorial and supergalactic planes are also shown.}
%


\begin{references}
\bibitem{bird95a}	D. J. Bird {\it et al.},
	Astrophys J. {\bf 441} (1995) 144.
\bibitem{hayashida94a}		N. Hayashida {\it et al.},
	Phys. Rev. Lett. {\bf 73} (1994) 3491.

\bibitem{mdl_object}	See, for example, 
	P. L. Biermann and P. A. Strittmatter, 
	Astrophys. J. {\bf 322} (1987) 643; 
	F. Takahara,
	Prog. Theor. Phys. {\bf 83} (1990) L1071; 
	J. P. Rachen and P. L. Biermann,
	Astron. Astrophys. {\bf 272} (1993) 161; 
	R. V. E. Lovelace, 
	Nature {\bf 262} (1976) 649.
\bibitem{mdl_relic}	See, for example, 
	V. Berezinsky, M. Kachelriess, and A. Vilenkin, 
	Phys. Rev. Lett. {\bf 79} (1997) 4302. 
\bibitem{mdl_td}	See, for example,
	C. T. Hill, D. N. Schramm and T. P. Walker,
	Phys. Rev. {\bf D36} (1987) 1007; 
	P. Bhattacharjee, C. T. Hill and D. N. Schramm,
	Phys. Rev. Lett. {\bf 69} (1992) 567.
\bibitem{mdl_grb}	See, for example,
	M. Vietri,
	Astrophys. J. {\bf 453} (1995) 883;
	E. Waxman,
	Phys. Rev. Lett. {\bf 75} (1995) 386.
\bibitem{berezinsky98a}	V. Berezinsky,
	hep$-$ph/9802351.

\bibitem{gzk66}
	K. Greisen,
	Phys. Rev. Lett. {\bf 16} (1966) 748; 
	G. T. Zatsepin and V. A. Kuz'min,
	Zh. Eksp. Teor. Fiz. {\bf 4} (1966) 114 
	[JETP Letters {\bf 4} (1966) 78].
\bibitem{hill85a}	C. T. Hill and D. N. Schramm,
	Phys. Rev. {\bf D31} (1985) 564.
\bibitem{berezinsky88a}	V. S. Berezinsky and S. I. Grigor'eva,
	Astron. Astrophys. {\bf 199} (1988) 1.
\bibitem{yoshida93a}	S. Yoshida and M. Teshima,
	Prog. Theor. Phys. {\bf 89} (1993) 833.
\bibitem{linsley86a}	J. Linsley, 
	J. Phys. G:Nucl. Part. Phys. {\bf 12} (1986) 51.
\bibitem{lawrence91a}	M. A. Lawrence, R. J. O. Reid and A. A. Watson, 
	J. Phys. G:Nucl. Part. Phys. {\bf 17} (1991) 733.
\bibitem{winn86a}	M. M. Winn {\it et al.},
	J. Phys. G:Nucl. Part. Phys. {\bf 12} (1986) 653.
\bibitem{efimov91a}	N. N. Efimov {\it et al.},
	in {\it Astrophysical Aspects of the Most Energetic Cosmic Rays}, 
	edited by M. Nagano and F. Takahara 
	(World Scientific, Singapore, 1991) p.20.
\bibitem{bird94a}	D. J. Bird {\it et al.},
	Astrophys J. {\bf 424} (1994) 491.
\bibitem{nagano92a}	M. Nagano {\it et al.},
	J. Phys. G: Nucl. Phys.  {\bf 18} (1992) 423.

\bibitem{chiba92a}	N. Chiba et al. ,
	Nucl. Instr. and Meth.  {\bf A 311} (1992) 338.
\bibitem{ohoka97a}	H. Ohoka {\it et al.},
	Nucl. Instr. and Meth. {\bf A 385} (1997) 268.
\bibitem{hillas71a}	A. M. Hillas {\it et al.},
	in {\it Proceedings of the 12th International Cosmic Ray Conference}, 
	Hobart, 1971, Vol. 3, p. 1001. 
\bibitem{yoshida94a}	S. Yoshida {\it et al.},
	J. Phys. G:Nucl. Part. Phys. {\bf 20} (1994) 651.
\bibitem{doi95a}	T. Doi {\it et al.},
	in {\it Proceedings of the 24th International Cosmic Ray Conference}, 
	Rome, 1995, Vol. 2, p. 764. 
\bibitem{sakaki98a}	N. Sakaki {\it et al.},
	{\it in preparation. }
\bibitem{dai88a}	H. Y. Dai {\it et al.},
	J. Phys. G: Nucl. Phys. {\bf 14} (1988) 793.
\bibitem{yoshida95a}	S. Yoshida et al. ,
	Astropart. Phys.  {\bf 3} (1995) 105. 

\bibitem{tibet96a}	M. Amenomori {\it et al.}, 
	Astrophys. J. {\bf 461} (1996) 408. 

\bibitem{thompson95a}	D. J. Thompson {\it et al.},
	Astrophys. J. Suppl. {\bf 101} (1995) 259.
\bibitem{veron83a}	M. P. Veron-Cetty and P. Veron, 
	Astron. Astrophys. Suppl. {\bf 53} (1983) 219.
\bibitem{hayashida96a}	N. Hayashida {\it et al.}, 
	Phys. Rev. Lett. {\bf 77} (1996) 1000.

\bibitem{teshima92a}	M. Teshima {\it et al.},
        Nucl. Phys. B (Proc. Suppl.) {\bf 28B} (1992) 169.
\bibitem{teshima94a}    M. Teshima, 
        in {\it Proceedings of the Tokyo Workshop on Techniques 
	for the Study of Extremely High Energy Cosmic Rays},
        edited by M. Nagano. (ICRR, University of Tokyo, 1993) p. 109.
\bibitem{hires94a}	E. Loh, 
        in {\it Proceedings of the Tokyo Workshop on Techniques 
	for the Study of Extremely High Energy Cosmic Rays},
        edited by M. Nagano. (ICRR, University of Tokyo, 1993) p. 105.
\bibitem{alseady97a}	M. Al-Seady {\it et al.},
        in {\it Proceedings of International Symposium on 
	Extremely High Energy Cosmic Rays: 
	Astrophysics and Future Observatories}, 
        edited by M. Nagano. (ICRR, University of Tokyo, 1996) p. 191.
\bibitem{cronin92a}	J. W. Cronin, 
	Nucl. Phys. B (Proc. Suppl.) {\bf 28B} (1992) 213.
\bibitem{auger97a}
        {\it THE PRIERRE AUGER OBSERVATORY DESIGN REPORT} (1997)

\end{references}
\end{document}